# On the  Limit of the Thermodynamic Stability  of Superheated Crystals and Mechanisms of Its Loss

V.I.Zubov

*Instituto de Física, Universidade Federal de Goiás, 74001-970, Goiânia, Go, Brazil*

*Department of Theoretical Physics, Peoples' Friendship University, Moscow, Russia*

E-mail: zubov@fis.ufg.br;  v_zubov@mail.ru

Using the correlative method of unsymmetrized self-consistent field for strongly anharmonic crystals, the thermodynamic stability and the mechanism its loss is studied for crystals with various types of the chemical bond. The following interaction potentials are utilized: the Lennard-Jones pairwise potential together with Axilrod - Teller three-body one for simple van der Waals crystals (solid Ar), effective interionic pairwise potential proposed by Schiff for a metal (Na) [Phys. Rev. **186**, 151 (1969)] and the Girifalco potential for fullerites ($C_{60}$) [J. Phys. Chem. **96**, 858 (1992)]. In the first and third cases, the FCC lattice becomes unstable because the isothermal bulk modulus $B_T \rightarrow 0$. For fullerites, the shearing coefficient $C_{44}$  goes to zero as well. Solid Na losses its stability when other shearing coefficient $C^T_{11} - C^T_{12}$ becomes zero.

С использованием корреляционного метода несимметризованного самосогласованного поля для сильно ангармонических кристаллов изучаются термодинамическая устойчивость и механизмы ее потери кристаллов с разными типами химической связи. Используются потенциалы взаимодействия: парный потенциал Леннард-Джонса вместе с трехчастичным потенциалом Аксильрода - Теллера для простых ван-дер-ваальсовых кристаллов (твердый Ar), эффективный межионный потенциал, предложенный Шиффом для металла (Na) и потенциал Жирифалко для фуллеритов ($C_{60}$). В первом и третьем случаях ГЦК решетка становится неустойчивой из-за того, что изотермический объемный модуль $B_T \rightarrow 0$. Для фуллеритов сдвиговый коэффициент $C_{44}$ также стремится к нулю. Твердый Na теряет устойчивость, когда другой сдвиговый коэффициент $C^T_{11} - C^T_{12}$ обращается в нуль.

The study of the thermodynamic stability, phase transitions, critical phenomena and metastable states is one of most important problems in modern statistical mechanics [1, 2]. It is well known that the thermodynamic equilibrium between two phases of a system along the transition curve is determined by the equality of their molar Gibbs free energies while the stability of each of them is determined by its minimum condition for a given phase. Disappearance of this minimum (its conversion into saddle point and then into the maximum) means a loss of thermodynamic stability by this phase (the spinodal). For the first-kind transitions, the equilibrium with other phase occurs early than the loss of the thermodynamic stability. Exceptions represent critical regions, which are observed at some transitions. In the general case, there exist metastable states. They correspond to local rather that absolute minimum of the Gibbs free energy and can take place under such conditions when another phase is the most stable possessing its absolute minimum [1, 2]. For crystals, the first-kind phase transitions are: sublimation, melting and polymorphic transformations.

Metastable states of fluid systems have been much studied [3]. Supercooling and superheating are common for polymorphic transitions [4]. Experimental studies of crystals superheated above their melting points have been hampered by occurrence of lattice defects, especially at surfaces, which are liable to be nuclei of the liquid phase. Well-prepared monocrystals can be superheated from the inside [5]. Metallic samples have to be superheated by high-powered current pulses [6]. It has been observed the laser-induced superheating of the Pb {111} surface considerably above its melting tmperature [7].

Theoretical investigations of metastable crystals are complicated by the strong anharmonicity of the lattice vibrations at temperatures close and especially above the melting points. The quasi-harmonic approximation leads to an estimation for the spinodal point of a crystal which is close or even below its melting temperature. However, a consistent inclusion of anharmonic terms [8 - 12] and computer simulations [13, 14] provides relative stability of crystals above their melting curves.

By this means, in spite of the above-mentioned experimental and theoretical difficulties, the existence of crystals superheated above their melting points has been firmly established. But the limit of the metastable region of such a crystal and the mechanism of the loss of its stability is a more complicated problem. Using the improved self-consistent phonon theory, Plakida and co-workers (e.g. [8]) have studied the stability of the face-centered cubic (fcc) lattice with nearest interactions described by the Morse potential [9] (see also [10]). Horner [15, 16] has noted that this theory cannot be used in its original form in the case of hard-core interactions, for instance, with the Lennard-Jones potential. To resolve this problem, he has proposed to utilize in addition a short-range correlation function.

The present paper is devoted to its analysis on the basis of the correlative method of unsymmetrized self-consistent field for strongly anharmonic crystals (CUSF).

The minimum conditions for the Gibbs free energy are positivity of the stability determinant, i.e. the Jakobian of a transformation from the intensive thermodynamic variables (the temperature $T$ and generalized forces $\{A\}$) to the extensive ones (the entropy $S$ and generalized coordinates $\{a\}$) together with its principal minors which are called the stability coefficients. In the case of crystals, $\{a\}$ are the components of the deformation tensor $\hat{\varepsilon}$ multiplied by the volume $V$ and $\{A\}$ those of the stress tensor $\hat{\sigma}$. For cubic crystals, this conditions take the form [17]

$$B_T > 0, \quad C_{11}^T > 0, \quad C_{11} - C_{12} > 0, \quad C_{44} > 0, \quad T/C_V > 0 \tag{1}$$

where $B_T$ is the isothermal bulk modulus, $C_{\alpha\beta}$ are the components of the elastic constant tensors (in the Foight notation) and $C_V$ is the isochoric specific heat. The violation of any of these inequality signifies the loss of stability (spinodal point).

In the CUSF [11. 18], the self-consistent potentials contain the main anharmonic terms, each of them being self-consistent with those of the power expansion of the initial potential energy. For the hard-core interatomic potentials, it provides the strong localization of each atom near the corresponding lattice point [18]. Because of this, the CUSF is free from the problem of the problem of hard-core correlations that has been discussed and resolved by Horner [15, 16] for the self-consistent phonon theory.

Usually, it is sufficient to include in the self-consistent potentials the anharmonicity up to the fourths order and to take into account the high-order terms by the perturbation theory. In this case, the Helmholtz free energy of a cubic one-component crystal with pairwise central interactions $\Phi(r)$ and its equation of state at the hydrostatic pressure $P$ are of the form, e.g. [11, 19]

$$F = N\left\{\frac{K_0}{2} - \frac{5\Theta}{24}\left(\frac{\beta}{X}\right)^2 - \frac{\Theta}{4}\left(X + \frac{5\beta}{6X}\right)^2 \right.$$
$$\left. - \Theta \ln\left[\left(\frac{3m^2\Theta^3}{\hbar^4 K_4}\right)^{3/4} D_{-15}\left(X + \frac{5\beta}{6X}\right)\right]\right\} + F^2 + F^H + F^Q; \tag{2}$$

$$P = -\frac{a}{3v}\left\{\frac{1}{2}\frac{dK_0}{da} + \frac{\beta\Theta}{2K_2}\frac{dK_2}{da} + \frac{(3-\beta)\Theta}{4K_4}\frac{dK_4}{da}\right\} + P^2 + P^H + P^Q. \tag{3}$$

Here $N$ is the number of molecules (Avogadro's number), $a$ is the lattice parameter or the nearest-neighbor distance, $v(a) = V/N$ the volume of the unit cell, $\Theta = kT$,

$$K_{2l} = \frac{1}{2l+1}\sum_{k\geq 1} Z_k \nabla^{2l}\Phi(R_k), \quad l = 0, 1, 2, \tag{4}$$

$Z_k$ and $R_k$ are the coordination numbers and radius, and $\beta(K_2(3/\Theta K_4)^{1/2})$ is an implicit function, determined by the transcendental equation

$$\beta = 3X\frac{D_{-25}(X + 5\beta/6X)}{D_{-15}(X + 5\beta/6X)}, \tag{5}$$

where $D_\nu$ are the parabolic cylinder functions.

In Eqs. (2) and (3), the expressions in the braces are the zeroth approximation that includes anharmonic terms up to the fourth order, $F^2$, $P^2$, $F^H$ and $P^H$ are the corrections by the perturbation theory which improve the zeroth approximation, in particular by taking into account the anharmonicity of higher orders, $F^Q$ and $P^Q$ the quantum corrections.

The CUSF enable one to take into account the many-body interactions, e.g. [11, 19] and the intramolecular degrees of freedom in molecular crystals [20]. It also has been generalized to crystals with compound lattices [21]. This method has been used to investigate strongly crystals with various types of chemical bonds: simple van der Waals crystals (the solidified noble gases), e.g. [11, 19, 21], ionic crystals (alkali halides) [22 - 24], a metal (solid sodium) [25] and high-temperature modification of fullerites($C_{60}$, $C_{70}$, $C_{76}$ and $C_{84}$) , e.g. [20, 26 - 28]. Results have been compared with available experimental data.

In all cases at temperatures bellow some limit $T_l$, the equation of state (3) has two roots $a_1(T)$ < $a_2(T)$, which coalesce at $T = T_l$. At higher temperatures it has no real solutions. The upper branches of isobars $a_2(T)$ correspond to the absolute unstable states, because the isothermal modulus $B_T(T, a_2)$ < 0. At $T \to T_l$, $B_T \to 0$ and the isobaric specific heat $C_P \to \infty$. Here we consider the thermodynamic stability of one-component crystals and the mechanism of its loss.

For Ar that has the FCC lattice, we use the Lennard-Jones pairwise potential

$$\Phi(r) = \varepsilon \left[ \left( r_0 / r \right)^{12} - 2 \left( r_0 / r \right)^6 \right], \tag{6}$$

with the minimum point $r_o = 3.780 \times 10^{-8}$ cm and the potential well depth $\varepsilon / k$ = 128.3 K [29]. The Axilrod - Teller three-body interactions [30]

$$\Phi_{(3)} \left( \vec{r}_i, \vec{r}_j, \vec{r}_k \right) = \frac{\nu}{\left( r_{ij} r_{jk} r_{ki} \right)^3} \left( 1 - 3 \cos \vartheta_i \cos \vartheta_j \cos \vartheta_k \right) \tag{7}$$

where $r_{ij}$, $r_{jk}$, $r_{ki}$ are the sides of the triangle formed be the centers of the three atoms and $\vartheta_i, \vartheta_j, \vartheta_l$ its interior angles are taken into account. For Na that possess the body-centered (BCC) lattice, we adopt the effective interionic pairwise potential proposed by Schiff [31],

$$\frac{\Phi(r)}{\varepsilon} = \left( A + \frac{B}{R^2} + \frac{C}{R^4} \right) \frac{\cos 2 k_F R}{R^3} + \left( D + \frac{E}{R^2} \right) \frac{\sin 2 k_F R}{R^4} , \tag{8}$$

the seven coordination spheres being included. Here $R = r/\sigma$, $\varepsilon/k$ = 599 K, $\sigma = 3.24 \times 10^{-8}$ cm. Other parameters lead to $r_o = 3.726$. Note that this potential provides the absolute thermodynamic stability of the BCC lattice and the relative stability of the FCC one. Finally, for the high-temperature modification of $C_{60}$ with the FCC lattice, we utilized the Girifalco potential [32],

$$\Phi_G(r) = -\alpha \left( \frac{1}{s(s-1)^3} + \frac{1}{s(s+1)^3} - \frac{2}{s^4} \right) + \beta \left( \frac{1}{s(s-1)^9} + \frac{1}{s(s+1)^9} - \frac{1}{s^{10}} \right), \tag{9}$$

in which $s = r/2a$ where $a$ is the radius of the hard core of a molecule which together with coefficients $\alpha$ and $\beta$ give $\varepsilon/k = 3218.4$ K and $r_0 = 10.0588 \times 10^{-8}$ cm.

In Fig. 1 we show the normal isobars $P = 1$ bar for solid Ar, Na and fullerite $C_{60}$ in the dimensionless form $T^* = \Theta/\varepsilon$, $a^* = a/r_0$. Symbols Ar, Na and $C_{60}$ signify the temperatures $T_l$ for them. They correspond to 118.3, 662.1 and 1917 K, respectively. Note that more realistic potential proposed for Ar by Barker and Bobetic [28] leads to similar results. Along the lower branches of the isobars the thermodynamic properties have been calculated including the coefficients of stability (1). In all cases, $T/C_V$ remain positive up to $T_l$. We shall analyze the behavior of other, mechanical stability coefficients.

The other figures demonstrate the mechanical stability coefficients: 1 - $C^T_{11}$, 2 - $B_T$, 3 - $C_{44}$, 4 - $C_{11}$ - $C_{12}$. For solid Ar (Fig.2), they remain positive up to $T_l$ where only $B_T$ goes to zero. Hence, for it (and for other simple van der Waals crystals), $T_S = T_l$, and near the spinodal point our estimation gives $B_T \sim (T_S - T)^{0.5}$. The BCC lattice of sodium (Fig. 3) losses its stability at $T = T_S \cong 498$ K $< T_l$ when the shearing coefficient $C_{11}$ - $C_{12}$ goes through, being $(C_{11} - C_{12}) \sim (T_S - T)$. In the case of $C_{60}$ at normal pressure (Fig. 4), $B_T$ goes to zero at $T_S = T_l$ together with other shearing coefficient $C_{44}$, both behavioring as $(T_S - T)^{0.5}$. Note that at high pressures, $C_{44}$ becomes zero at $T_S(P) < T_l(P)$ with positive $B_T$. In this case, $C_{44} \sim (T_S - T)$.

So, the thermodynamic stability of a crystal and the mechanism of its loss are due to the nature of its interaction forces and type of the crystalline lattice which also depends on interactions. The mechanism of the loss can change with pressure. More thorough analysis of such dependence on peculiarities of interatomic potentials in the future should be of great interest.

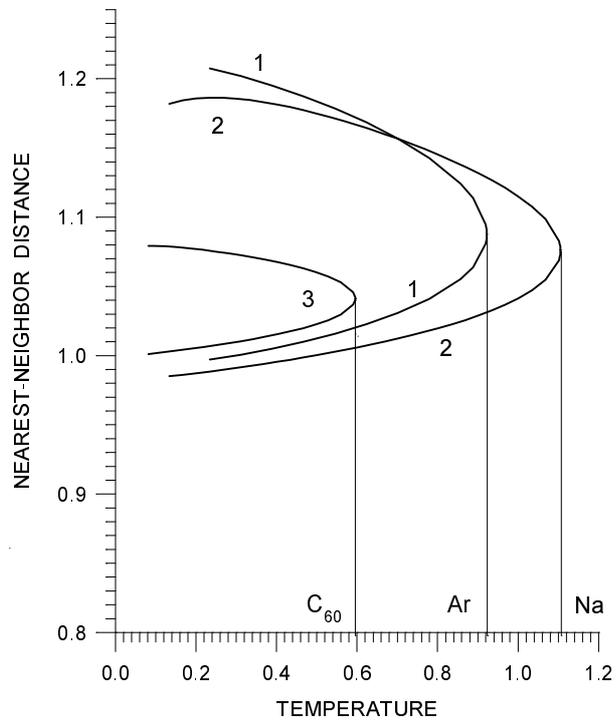

Fig. 1. The normal isobars of solid Ar (1), Na (2) and $C_{60}$ fullerite (3), in the dimensionless form $T^* = \Theta/\varepsilon$ and $a^* = a/r_0$ .

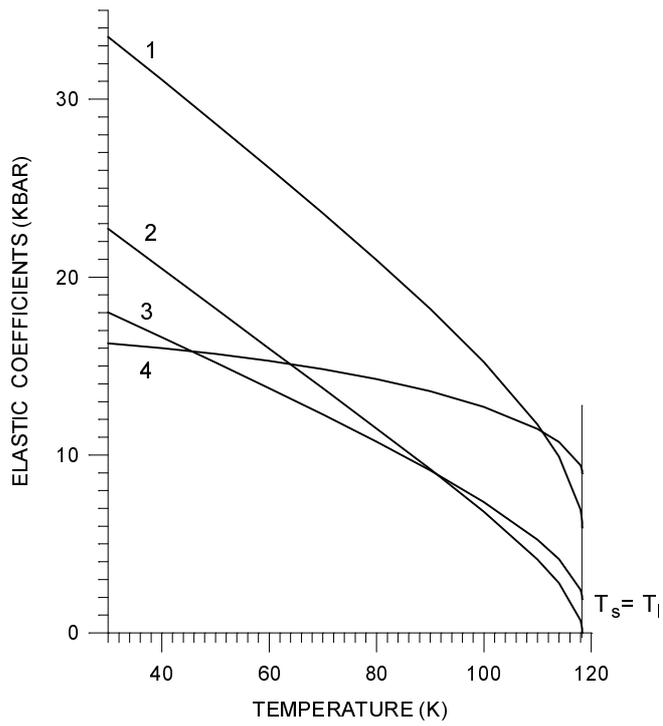

Fig. 2. Stability coefficients $C^T_{11}$ (1), $B_T$ (2), $C_{44}$ (3) and $C_{11} - C_{12}$ (4) of solid Ar.

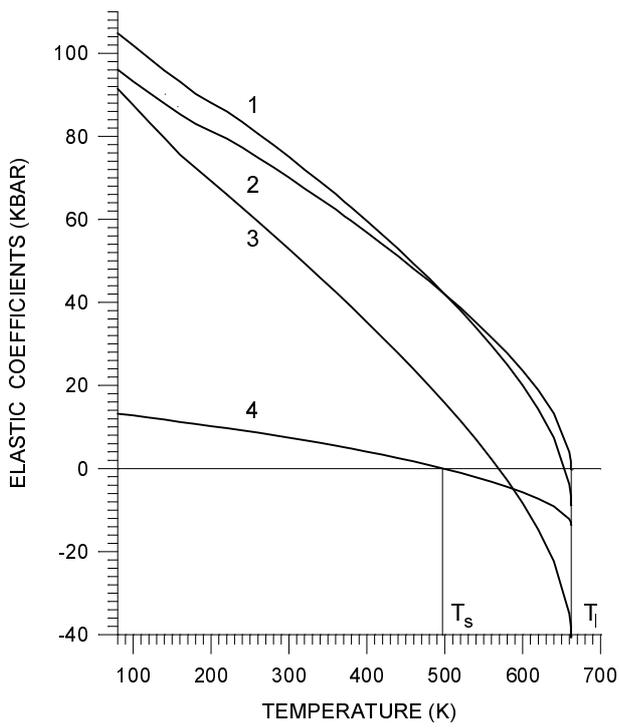

Fig. 3. Stability coefficients $C^{T}_{11}$ (1), $B_{T}$ (2), $C_{44}$ (3) and $C_{11} - C_{12}$ (4) of solid Na.

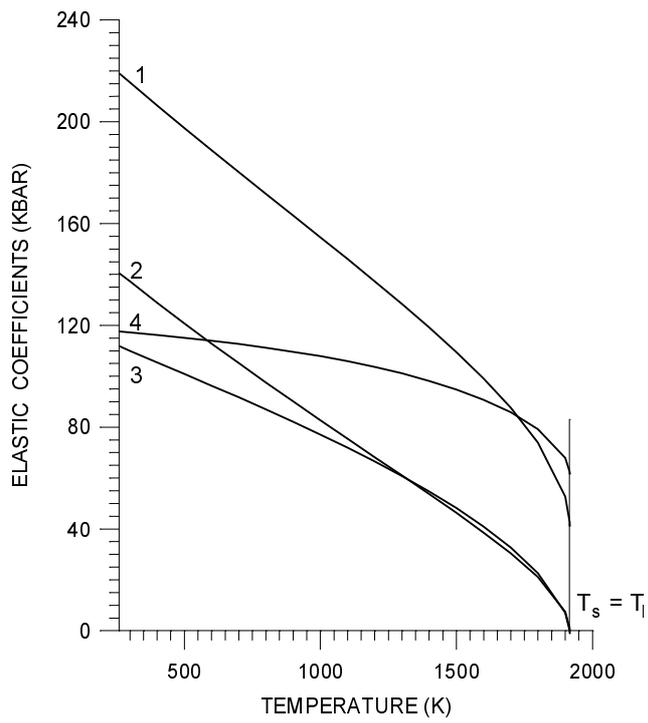

Fig. 4. Stability coefficients $C^{T}_{11}$ (1), $B_T$ (2), $C_{44}$ (3) and $C_{11} - C_{12}$ (4) of $C_{60}$ fullerite.